\documentclass[prl,amsmath,superscriptaddress,twocolumn,showpacs]{revtex4}
\usepackage{bm}
\usepackage{amssymb}
\usepackage{graphicx}

\begin{document}

\title{Steps and dips in the ac conductance and noise of mesoscopic structures}

\author{O. Entin-Wohlman}
\email{entin@post.tau.ac.il}

\affiliation{Department of Physics and the Ilse Katz Center for
Meso- and Nano-Scale Science and Technology, Ben Gurion
University, Beer Sheva 84105, Israel}

\affiliation{Albert Einstein Minerva Center for Theoretical
Physics, Weizmann Institute of Science, Rehovot 76100, Israel}

\author{Y. Imry}

\affiliation{Department of Condensed Matter Physics,  Weizmann
Institute of Science, Rehovot 76100, Israel}

\author{S. A. Gurvitz}

\affiliation{Department of Particle Physics,  Weizmann Institute
of Science, Rehovot 76100, Israel}

\author{A. Aharony}

\affiliation{Department of Physics and the Ilse Katz Center for
Meso- and Nano-Scale Science and Technology, Ben Gurion
University, Beer Sheva 84105, Israel}

\date{\today}

\pacs{73.22.Gk,72.15.Qm,73.21.La,73.23.Hk}

\begin{abstract}

The frequency dependence of the equilibrium ac conductance (or the
noise power spectrum) through a mesoscopic structure is shown to
exhibit steps and dips. The steps, at energies related to the
resonances of the structure, are closely related to the partial
Friedel phases of these resonances, thus allowing a direct
measurement of these phases (without interferometry). The dips in
the spectrum are related to a destructive interference in the
absorption of energy by transitions between these resonances, in
some similarity with the Fano effect.

\end{abstract}

\maketitle

 In recent years, it is becoming clear  that measurements of the
noise power spectrum of a complex mesoscopic structure,
$C(\omega)$, can provide invaluable information on its physics
\cite{book,Butt-Blant}. Examples include the information on the
transmission eigrnvalues \cite{Been-Butt}, and the effective
charge of the quasiparticles, provided by shot-noise measurements
\cite{charge}. The noise spectrum is proportional to the
emission-absorption spectrum of the system \cite{abs}, which is
related \cite{linresp} to its ac conductivity, ${\cal G}(\omega)$.
So far, much of the information was obtained at rather low
frequencies (with important exceptions, \cite{Glattli06}).
However, it is clear that much further dynamic information will
follow from higher-frequency measurements, which we study in the
present Letter. Some of the motivation for this work arose from
the report of a structure at the Larmor frequency in the power
spectrum of a single magnetic moment on a surface, measured at
high frequencies with the STM technique \cite{manassen}.

 The {\em linear} ac
conductance, ${\cal G}(\omega )$, is determined \cite{linresp} by
the equilibrium properties of the un-biased system. It is related
to the equilibrium value of the noise power spectrum via the
fluctuation-dissipation theorem \cite{linresp},
\begin{align}
C(\omega )&=\frac{2\omega }{e^{\beta\omega}-1}\Re\Bigl ({\cal
G}(\omega )\Bigr )\ ,
\label{FD}
\end{align}
where $\beta=1/kT$ ($T$ is the temperature), energies are measured
from the Fermi energy ($\epsilon_F=0$) and $C(\omega )$ is the
(unsymmetrized) current-current correlation function
\begin{align}
C^{}_{}(\omega )=\int dt e^{-i\omega t}\langle
\delta\hat{I}^{}_{}(t)  \delta\hat{I}_{}(0)\rangle \
,\label{noise}
\end{align}
with
$\delta\hat{I}^{}_{}=\hat{I}^{}_{}-\langle\hat{I}^{}_{}\rangle $,
and $\hat{I}$ is the net current operator through the system (we
use $\hbar =1$).

As we show below, under appropriate conditions the frequency
dependence of  $C(\omega)$ is determined by the energy dependence
of the fundamental Friedel phase, $\delta(\epsilon)$, which
relates to the charge accumulated in the region of the mesoscopic
structure. In particular, as $\omega$ crosses a resonance energy,
$C(\omega)$ follows the increase of $\delta(\omega)$ by $\pi$. We
thus suggest that $\delta$ can be deduced from measurements of
$C(\omega)$ or of ${\cal G}(\omega )$. Except for special points,
where the transmission vanishes, $\delta(\epsilon)$ is related
\cite{yeyati} to the transmission phase  of the quantum scattering
through the mesoscopic structure, whose measurement using the
Aharonov-Bohm interferometer has attracted much discussion
\cite{EXP}. Here we propose an alternative method to measure this
phase.

When the mesoscopic structure has more than one resonance,  we
sometimes find  dips in $C(\omega)$, when $\omega$ is close to the
difference between the energies of these resonances. Since
$C(\omega)$ is directly related to the absorption of energy by the
system at energy $\omega$, such dips must arise from a destructive
interference between the quantum amplitudes for the transitions
involving these resonances, in some analogy to the Fano effect
\cite{Fano}.

 In the absence of interactions, ${\cal
G}(\omega )$ and $C(\omega)$  of a mesoscopic system can be
conveniently described by its (energy-dependent) scattering
matrix, $S(\epsilon )$. For clarity we concentrate on systems
connected to electronic reservoirs by two single-channel leads.
Then
\begin{align}
S(\epsilon )&=\left [\begin{array}{cc}r(\epsilon )&t(\epsilon)\\
t(\epsilon )&r'(\epsilon )\end{array}\right ]\nonumber\\
&\equiv e^{i\delta (\epsilon )}\left [\begin{array}{cc}\cos\theta
(\epsilon )e^{i\alpha (\epsilon )}&i\sin\theta(\epsilon)
\\
i\sin\theta (\epsilon) &\cos\theta (\epsilon )e^{-i\alpha
(\epsilon )}\end{array}\right ]\ .\label{SCATM}
\end{align}
Here,  $r(\epsilon )$ and $r'(\epsilon )$ are the reflection
amplitudes of the structure, and $t(\epsilon )$ is its
transmission amplitude. [Without magnetic fields, the system
possesses time-reversal symmetry and hence $t(\epsilon
)=t'(\epsilon )$.] The second equality in Eq. (\ref{SCATM})
depicts $S(\epsilon)$ in its parametric form, in which the phase
$\alpha$ represents deviations from left-right symmetry, which
result in $r'\neq r$, and $\delta (\epsilon )$ is the Friedel
phase.  One may find $C(\omega )$ in terms of the scattering
matrix elements either by employing the Kubo linear response
theory to calculate ${\cal G}(\omega )$  or by using the
scattering formalism \cite{butti,levinson} to find $C(\omega )$
directly. The result is
\begin{align}
C(\omega )=\frac{e^{2}}{4\pi }\int &d\epsilon f(\epsilon +\omega
)\bigl (1-f(\epsilon )\bigr )  {\cal C}(\epsilon ,\omega )\
.\label{C}
\end{align}
where $f(\epsilon )=(e^{\beta\epsilon}+1)^{-1}$ is the Fermi
function and
\begin{align}
&{\cal C}(\epsilon ,\omega  )=
 {\Re}\Bigl (2 +2t(\epsilon +\omega
)t^{\ast}(\epsilon ) \nonumber\\
&-r(\epsilon +\omega )r^{\ast}(\epsilon )-r'(\epsilon +\omega
)r'^{\ast}(\epsilon )\Bigr )\ ,\label{CW}
\end{align}
In the rest of this paper we present explicit results only at
$T=0$. In that case, the integration is over the range
$0<\epsilon<-\omega$, and therefore $C(\omega)\ne 0$ only for
$\omega<0$.

Consider first  a single localized level, of energy
$\epsilon^{}_{d}$, coupled to the left and the right leads by the
tunnelling matrix elements $V^{}_{L}$ and $V^{}_{R}$ \cite{COM1},
respectively. In this case,
\begin{align}
S(\epsilon )&=-1+\frac{2\pi i{\cal N}}{\epsilon -\epsilon^{}_{d}+i
\Gamma}\left [\begin{array}{cc}V^{2}_{L}&V^{}_{L}V^{}_{R}\\
V^{}_{L}V^{}_{R}&V^{2}_{R}\end{array}\right ]\ ,\label{S_single}
\end{align}
where ${\cal N}$ is the density of states at the Fermi energy of
the leads, and $\Gamma =\pi {\cal N}(V^{2}_{L}+V^{2}_{R})$ is the
resonance width. The Friedel phase is given by $\Gamma {\rm
cot}\delta (\epsilon )=\epsilon^{}_{d}-\epsilon$, such that
$\delta $ decreases from $\pi$ to zero as $\epsilon$ increases
from $-\infty$ to $\infty$, passing through $\pi /2$ at resonance.
When this system has left-right symmetry, $V^{}_{L}=\pm V^{}_{R}$,
$S(\epsilon)$ is completely determined by $\delta (\epsilon )$
alone, i.e., in the notations of Eq. (\ref{SCATM}) $\alpha =0$,
$\cos\theta =-\cos\delta$, and $\sin\theta =\pm \sin \delta$. Then
${\cal C}(\epsilon ,\omega )= 2(\sin^{2}[\delta (\epsilon
)]+\sin^{2}[\delta (\epsilon +\omega )])$. [Note that for a
symmetric Breit-Wigner resonance one has $|t|^2=\sin^{2}\delta$.]
Consequently, at $T=0$,
\begin{align}
C(\omega )&=\frac{e^{2}}{2\pi}\int_{\omega }^{-\omega }d\epsilon
\sin^{2}[\delta(\epsilon )]\ ,\ \omega\leq 0\ ,\label{CT}
\end{align}
and $C$ is a monotonic increasing function of $|\omega|$, growing
over a region of width $\Gamma$ near $\epsilon \sim
\epsilon^{}_d$, where the integrand is large. Moreover, since
$\Gamma d\delta =\sin^{2}(\delta )d\epsilon$, one has
\begin{align}
C(\omega )&=\frac{e^{2}\Gamma}{2\pi}[\delta (-\omega )-\delta
(\omega )]\ ,\ \omega\leq 0\ ,\label{Cdelta}
\end{align}
and thus $2 \pi C(\omega)/(e^2 \Gamma)$ follows exactly the growth
of the Friedel phase from zero to $\pi$. Explicit calculations
show that this qualitative step-like behavior also appears for the
non-symmetric case, when $V^{}_L /V^{}_R \ne \pm 1$.

A far more intricate behavior is obtained for a system of two
localized levels (at energies $\epsilon^{}_{1}$ and
$\epsilon^{}_{2}$), connected in parallel to the two leads
\cite{COM2},via tunnelling matrix elements $V^{}_{L1}$ and
$V^{}_{L2}$ (for the left) and $V^{}_{R1}$ and $V^{}_{R2}$ (for
the right lead) \cite{COM1}. The resonance widths of the two
separate levels are $\Gamma^{}_{i}=\pi{\cal
N}(V^{2}_{Li}+V^{2}_{Ri})$ for $i=1,2$. For this structure, the
solution of the scattering problem yields
\begin{widetext}
\begin{align}
S(\epsilon )=-1+\frac{2\pi i{\cal N}}{D(\epsilon )}\left
[\begin{array}{cc}V^{2}_{L1}(\epsilon
-\epsilon^{}_{2})+V^{2}_{L2}(\epsilon
-\epsilon^{}_{1})+i\frac{\Gamma^{}_{1}\Gamma^{}_{2}-\Gamma^{2}_{12}}{\pi{\cal
N}}&V^{}_{L1}V^{}_{R1}(\epsilon
-\epsilon^{}_{2})+V^{}_{L2}V^{}_{R2}(\epsilon -\epsilon^{}_{1})\\
V^{}_{L1}V^{}_{R1}(\epsilon
-\epsilon^{}_{2})+V^{}_{L2}V^{}_{R2}(\epsilon -\epsilon^{}_{1})&
V^{2}_{R1}(\epsilon -\epsilon^{}_{2})+V^{2}_{R2}(\epsilon
-\epsilon^{}_{1})+i\frac{\Gamma^{}_{1}\Gamma^{}_{2}-\Gamma^{2}_{12}}{\pi{\cal
N}}\end{array}\right ]\ ,
\end{align}
\end{widetext}
where $D(\epsilon )=(\epsilon
-\epsilon^{}_{1}+i\Gamma^{}_{1})(\epsilon
-\epsilon^{}_{2}+i\Gamma^{}_{2})+\Gamma^{2}_{12}$
 and
$\Gamma^{}_{12}=\pi{\cal
N}(V^{}_{L1}V^{}_{L2}+V^{}_{R1}V^{}_{R2})$.

 We next present a few
typical graphs, calculated using these exact expressions at $T=0$.
We then explain some of the observed features using approximate
expressions. Figure \ref{NOISE2} shows $2C(\omega)/e^2$ for two
resonances which are located below the Fermi energy. All the
graphs exhibit two steps in $2C(\omega)/e^2$, from 0 to
$\Gamma^{}_1$ and then from there to $\Gamma^{}_1+\Gamma^{}_2$.
The small differences between the graphs are due to the magnitude
of the left-right asymmetry. For sufficiently separated
resonances, one should not be surprised to see that each resonance
is indeed described by Eq. (\ref{Cdelta}), thus capturing the
behavior of the partial Friedel phases for each resonance. The
same curves are found when both resonances are above the Fermi
level.

\begin{figure}
\includegraphics[width=6cm]{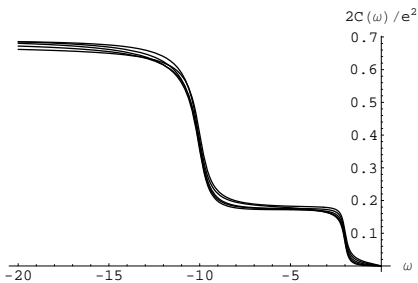}
\caption{The $T=0$ noise spectrum of two localized levels located
below the Fermi energy, ($\epsilon^{}_{1}=-2$ and
$\epsilon^{}_{2}=-10$), of widths $\Gamma^{}_{1}=0.18$ and
$\Gamma^{}_{2}=0.5$. The different curves correspond to different
values of the ratio $V^{}_{L2}/V^{}_{R2}$, with values
$1,~0.3,~0,~-0.3,~-1$ (keeping $\Gamma^{}_2$ fixed and
$V^{}_{L1}=V^{}_{R1}$). } \label{NOISE2}
\end{figure}

A surprisingly different result appears when the  two resonances
are located on both sides of the Fermi level, allowing absorption
between them. Figure \ref{NOISE} shows $2C(\omega)/e^2$ for this
case, for different ratios $V^{}_{L2}/V^{}_{R2}$. For
$V^{}_{L2}=V^{}_{R2}$, one finds the same monotonic two-step noise
as found in Fig. \ref{NOISE2}. However, as this ratio decreases,
there appears an increasing novel dip in the noise, at $|\omega|$
close to the difference $\epsilon^{}_1-\epsilon^{}_2$.

\begin{figure}
\includegraphics[width=6cm]{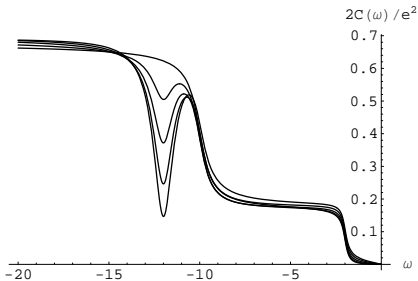}
\caption{Same as Fig. \ref{NOISE2}, but with $\epsilon^{}_{1}=2$
and $\epsilon^{}_{2}=-10$. The dip near $\omega=-12$ increases as
the ratio $V^{}_{L2}/V^{}_{R2}$ decreases from $1$ to $-1$, via
$0.3,~0,~-0.3$. } \label{NOISE}
\end{figure}

Most of the observed features can be understood from Eq.
(\ref{CW}): away from resonances, $t(\epsilon)$ is small and
$r(\epsilon)$ and $r'(\epsilon)$ are close to $-1$, so that ${\cal
C}(\epsilon,\omega)$ is small. When $\epsilon$ crosses one of the
resonances (say $\epsilon^{}_1$) then $r(\epsilon)$ and
$r'(\epsilon)$ are both small, the last two terms in Eq.
(\ref{CW}) are small, and the integrand ${\cal
C}(\epsilon,\omega)$ is dominated by
$2\Re[1+t(\epsilon+\omega)t^\ast(\epsilon)]$. When
$\epsilon+\omega$ is not near the other resonance, then
$t(\epsilon+\omega)$ is also small, and ${\cal C}$ is of order 2,
giving a large contribution to the integral. This yields the two
steps in $C(\omega)$. The only potential exception to this can
happen when $\epsilon+\omega$ is also near a resonance, when
$|t(\epsilon)|$ and $|t(\epsilon+\omega)|$ are {\em both} of order
unity. In this case, the result depends on the magnitude and phase
of $t(\epsilon+\omega)t^\ast(\epsilon)$. For well separated
resonances, the behavior of $t(\epsilon)$ near each of them can be
approximated by Eq. (\ref{S_single}), namely $t(\epsilon) \sim
V^{}_L V^{}_R$. Thus,
$t(\epsilon^{}_1)t^\ast(\epsilon^{}_1+\omega) \sim
V^{}_{L1}V^{}_{R1}V^{}_{L2}V^{}_{R2}$, and the destructive
interference in ${\cal C}$ is largest when
$V^{}_{L1}V^{}_{R1}V^{}_{L2}V^{}_{R2}=-\Gamma^{}_1\Gamma^{}_2/4$.

 In fact, the contribution of a scattering state $|\epsilon\rangle$ at
energy $\epsilon$ to Eq. (\ref{noise}) can be written as the
probability of the absorption of energy $\omega$ by a transition
between $|\epsilon\rangle$ and $|\epsilon+\omega\rangle$,
$|\langle\epsilon|\hat I|\epsilon+\omega\rangle|^2$. Consider the
case when both $|\epsilon\rangle$ and $|\epsilon+\omega\rangle$
come from the left.  Each of these contains an incoming, a
transmitted and a reflected waves. One then ends up with
$\langle\epsilon|\hat I|\epsilon+\omega\rangle \propto
[1+t(\epsilon+\omega)t^\ast(\epsilon)-r(\epsilon+\omega)r^\ast(\epsilon)]$.
The reduction of ${\cal C}$ when both $\epsilon$ and
$\epsilon+\omega$ hit the two resonances is then a consequence of
destructive interference between the first two terms here, in some
similarity to  the original Fano effect \cite{Fano}. Whether the
resulting dip is large or small thus depends on the relative
phases of $t(\epsilon^{}_1)$ and $t(\epsilon^{}_2)$ (see above)
and on the overall weight of this term in the final integral. This
weight is large only when the two resonances are on the two sides
of the Fermi energy.

A more quantitative insight into these results can be achieved by
looking at approximate analytical expressions, derived when the
resonance locations are well-separated, such that
$(\Gamma^{}_{1}-\Gamma^{}_{2})^{2},~ 4\Gamma^{2}_{12}\ll
(\epsilon^{}_{1}-\epsilon^{}_{2})^{2}$. Although the resonance
locations and their respective widths are modified
($\epsilon^{}_{1,2} \to \epsilon^{}_{a,b}$, $\Gamma^{}_{1,2} \to
\Gamma^{}_{a,b}$) once the two levels are connected to the leads
to form a `ring', in this limit these modifications are small.
Writing 
$D(\epsilon )=(\epsilon -\epsilon^{}_{a}+i\Gamma^{}_{a})(\epsilon
-\epsilon^{}_{b}+i\Gamma^{}_{b})$,
 one finds that 
up to  order  $\Gamma^{2}_{12}$
the resonance widths are unchanged, $\Gamma^{}_{a}\simeq
\Gamma^{}_{1}$ and $\Gamma^{}_{b}\simeq\Gamma^{}_{2}$, while
$\epsilon^{}_{a,b}\simeq (\epsilon^{}_{1}+\epsilon^{}_{2}\pm\Omega
)/2$,
with the modified energy difference between the two resonances
given by $\Omega^{2}\simeq
(\epsilon^{}_{1}-\epsilon^{}_{2})^{2}-4\Gamma^{2}_{12}>0$. The
Friedel phase of the combined structure is now given by $\delta
(\epsilon )=\delta^{}_{a}(\epsilon )+\delta^{}_{b}(\epsilon )$,
where the `partial' Friedel phases of the two resonances,
$\delta^{}_{a}$ and $\delta^{}_{b}$, are given by
\begin{align}
\Gamma^{}_{a}{\rm cot}[\delta^{}_{a}(\epsilon
)]=\epsilon^{}_{a}-\epsilon\ ,\ \ \ \
\Gamma^{}_{b}{\rm cot}[\delta^{}_{b}(\epsilon
)]=\epsilon^{}_{b}-\epsilon\ .\label{PARTIALF}
\end{align}

The analytic expressions turn out to be much simpler for the two
extreme cases of the left-right symmetry, namely for
$V^{}_{L1}=\pm V^{}_{R1}$ and $V^{}_{L2}=\pm V^{}_{R2}$, i.e. when
$r(\epsilon )=r'(\epsilon )$. Had we further assumed that
$V^{}_{L1}=V^{}_{R1}$ and $V^{}_{L2}=V^{}_{R2}$ \cite{COM3}, we
would have found
\begin{align}
r&=-e^{i(\delta^{}_{a}+\delta^{}_{b})}\cos[\delta^{}_{a}+\delta^{}_{b}]\
,\nonumber\\
t&=-ie^{i(\delta^{}_{a}+\delta^{}_{b})}\sin[\delta^{}_{a}+\delta^{}_{b}]\
.\label{SIMPLE}
\end{align}
Namely, the phase $\theta$ in
Eq. (\ref{SCATM}) is equal (up to an irrelevant sign) to the total
Friedel phase, $\delta(\epsilon)$. Therefore, at $T=0$,
$C(\omega)$ is given by Eq. (\ref{CT}), with $\delta
=\delta^{}_{a}+\delta^{}_{b}$. It is a monotonic function of the
frequency, as indeed found for this symmetric case in the top
curve of Fig. \ref{NOISE} and in Fig. \ref{NOISE2}. The two steps
in this curve indeed capture the energy dependence of
$\delta^{}_a$ and of $\delta^{}_b$.

An entirely different picture is found  when $V^{}_{L1}=
V^{}_{R1}$ and $V^{}_{L2}=- V^{}_{R2}$, corresponding to the
lowest curve in Fig. \ref{NOISE}. Then the same approximations
yield
\begin{align}
r&=-e^{i(\delta^{}_{a}+\delta^{}_{b})}\cos[\delta^{}_{a}-\delta^{}_{b}]\
 ,\nonumber\\
 t&=-ie^{i(\delta^{}_{a}+\delta^{}_{b})}\sin[\delta^{}_{a}-\delta^{}_{b}]\
 ,
\end{align}
and the phase $\theta$ is given by the {\em difference} between
the two partial Friedel phases (up to an irrelevant sign).

To explain analytically the difference between the upper and lower
curves in Fig. \ref{NOISE}, we take the additional assumption
$\Gamma^{}_{1}=\Gamma^{}_{2}\equiv\Gamma /2$. In that case,  Eq.
(\ref{CW}) becomes ${\cal C}(\epsilon ,\omega )= 2(1-\cos [\theta
(\epsilon ) +\theta (\epsilon +\omega )]\cos [\delta (\epsilon
+\omega )-\delta (\epsilon )])= 2(\sin^{2}[\delta^{}_{a}(\epsilon
)-\delta^{}_{b}(\epsilon +\omega )]
+\sin^{2}[\delta^{}_{b}(\epsilon )-\delta^{}_{a}(\epsilon +\omega
)])$. Using the definitions of $\delta^{}_a$ and $\delta^{}_b$,
Eqs. (\ref{PARTIALF}), one finds
\begin{align}
{\cal C}(\epsilon ,\omega )=&\frac{8(\Omega +\omega
)^{2}}{\Gamma^{2}}\sin^{2}[\delta^{}_{a}(\epsilon
)]\sin^{2}[\delta^{}_{b}(\epsilon +\omega )]\nonumber\\
&+\frac{8(\Omega -\omega
)^{2}}{\Gamma^{2}}\sin^{2}[\delta^{}_{b}(\epsilon
)]\sin^{2}[\delta^{}_{a}(\epsilon +\omega )]\ .
\end{align}
The $\epsilon-$integration of this function, yielding $C(\omega
)$,
can be performed
straightforwardly (for $T=0$), as done in the figures. This
results with two complicated terms, which multiply $F(\omega)$ and
$F(-\omega)$, where
\begin{align}
F(\omega )=\frac{(\Omega +\omega )^{2}}{\Gamma^{2}+(\Omega +\omega
)^{2}}\ .\label{F}
\end{align}
However, it is instructive to further expand these terms to
leading order in the width $\Gamma$. For $\omega<0$ this yields
\begin{widetext}
\begin{align}
\frac{2\pi}{e^{2}}C(\omega )&\simeq F(\omega )\Bigl
(\int_{0}^{-\omega}d\epsilon\sin^{2}[\delta^{}_{a}(\epsilon
)]+\int_{\omega}^{0}d\epsilon\sin^{2}[\delta^{}_{b}(\epsilon
)]\Bigr )+F(-\omega )\Bigl
(\int_{0}^{-\omega}d\epsilon\sin^{2}[\delta^{}_{b}(\epsilon
)]+\int_{\omega}^{0}d\epsilon\sin^{2}[\delta^{}_{a}(\epsilon
)]\Bigr )\nonumber\\
=&\frac{\Gamma}{2} F(\omega )\Bigl (\delta^{}_{a}(-\omega
)-\delta^{}_{a}(0)+\delta^{}_{b}(0)-\delta^{}_{b}(\omega )\Bigr )+
\frac{\Gamma}{2}F(-\omega )\Bigl (\delta^{}_{b}(-\omega
)-\delta^{}_{b}(0)+\delta^{}_{a}(0)-\delta^{}_{a}(\omega )\Bigr )\
 .\label{C2L}
\end{align}
\end{widetext}
This approximate expression shows that (i) for $|\Omega+\omega|
\lesssim \Gamma$, the noise spectrum  follows closely the
monotonic frequency-dependence of the partial Friedel phases
$\delta^{}_{a}$ and $\delta^{}_{b}$,  and (ii) the dip results
from the function $F$.

Let us  examine the $\omega$-dependence in (\ref{C2L}). To this
end we note that $\delta^{}_{a}(-\omega )-\delta^{}_{a}(0)$
differs significantly from zero once $0<\epsilon^{}_{a}<-\omega $,
while $\delta^{}_{a}(0 )-\delta^{}_{a}(\omega )$ mainly
contributes when $\omega <\epsilon^{}_{a}<0$ (and similarly for
 $\delta^{}_{b}$). Suppose that the two levels are
located on both sides of the Fermi level, such that
$\epsilon^{}_{a}>0>\epsilon^{}_{b}$ [and hence
 $\Omega =\epsilon_a-\epsilon_b
>\epsilon^{}_{a}$].
Then the phase differences in the second term of Eq. (\ref{C2L})
are rather small, whereas those in the first term are substantive.
As $|\omega |$ increases, firstly
$\delta^{}_{b}(0)-\delta^{}_{b}(\omega )$ comes into play, giving
the first step of the curve at about $\omega \simeq
\epsilon^{}_{b}$, and as $|\omega |$ increases further
$\delta^{}_{a}(-\omega )-\delta^{}_{a}(0)$ joins in and yields the
second step at about $|\omega |\simeq\epsilon^{}_{a}$ (see Fig.
\ref{NOISE}). As $|\omega|$ increases further, the function
$F(\omega )$, Eq. (\ref{F}), vanishes at $|\omega|=\Omega$,
resulting in the pronounced dip. In contrast, when the two levels
are both on the same side of the Fermi energy, e.g.,
$\epsilon^{}_{a}>\epsilon^{}_{b}>0$ (in which case
$\epsilon^{}_{a}>\Omega$) one encounters the two steps as in the
case discussed above, but there will be no dip, since the
contribution from the first term in Eq. (\ref{C2L}) becomes
effective only for $|\omega |\simeq\epsilon^{}_{a}$, namely, at
absolute frequency {\em larger} than $\Omega$ (see Fig.
\ref{NOISE2}). As Fig. \ref{NOISE} shows, all of these features
survive whenever $V^{}_{L2}/V^{}_{R2}\ne 1$, namely when
$\theta(\epsilon)$ deviates from $\delta(\epsilon)$.

So far, we have discussed the `ring' of two localized levels in
the absence of a magnetic flux. Interestingly, a flux through the
ring modifies the product $V^{}_{L1}V^{}_{L2}V^{}_{R1}V^{}_{R2}$
by the Aharonov-Bohm phase factor, $e^{i\phi}$, where $\phi$ is
proportional to the flux threading the ring. It is interesting to
note that as $\phi$ changes by $\pi$, the sign of this product
changes. We thus expect that the new dip will appear and disappear
periodically as the flux increases.

It should be pointed out that our qualitative results remain valid
also when one applies a small potential difference across the
system. The power spectrum of the noise of biased systems depends
on the (different) Fermi functions of the two leads
\cite{butti,levinson}. As long as the bias is not too large, it
will not affect much the negative-frequency part of the spectrum.
However, a finite bias does induce a small positive-frequency
part, which may also contain steps and dips. Our results for a
general bias will be reported separately.

This work was supported by a Center of Excellence of the Israel
Science Foundation, by a grant from by the German Federal Ministry
of Eduction and Research (BMBF) within the framework of the
German-Israeli Project Cooperation (DIP), and from the US-Israel
Binational Foundation (BSF). We also acknowledge the hospitality
of the PITP at UBC (OEW, YI and AA) and at LANL (SAG).

\date{\today}

\maketitle


\begin{thebibliography}{999}
\bibitem{COM1}
The tunnelling matrix elements can be chosen to be real since the
system posseses time-reversal symmetry.
\bibitem{COM2}

In fact, any mesoscopic system with $N$ inter-connected localized
levels can be turned into a system of $N$ levels coupled in
parallel to the two leads. One just has to diagonalize the
decoupled system Hamiltonian, and then re-define  the tunneling
matrix elements connecting the parallel levels to the leads.
\bibitem{COM3}

In fact, for this symmetric case Eq. (\ref{SIMPLE}) is valid
whenever $V^{}_{L1}V^{}_{R1}V^{}_{L2}V^{}_{R2}>0$. In these
equations, the phase prefactors are exact, while the trigonometric
factors are correct to leading order in the widths.
\end{thebibliography}
\end{document}